\documentstyle[preprint,aps]{revtex}
\begin{document}
\draft
\title{Hierarchical Structure in Financial Markets}
 
\author{Rosario N. Mantegna}

\address{
Istituto Nazionale per la Fisica della Materia, Unit\`a di Palermo\\
and\\
Dipartimento di Energetica ed Applicazioni di Fisica,
Universit\`a di Palermo, Viale delle Scienze, I-90128,
Palermo, Italy}

 
\maketitle
 
\begin{abstract}
I find a topological arrangement of stocks traded in a financial market
which has associated a meaningful economic taxonomy. The topological space
is a graph connecting the stocks of the portfolio analyzed. The graph is 
obtained starting from the matrix of correlation coefficient computed between 
all pairs of stocks of the portfolio by considering the synchronous time evolution 
of the difference of the logarithm of daily stock price. The hierarchical 
tree of the subdominant ultrametric space associated with the graph provides 
information useful to investigate the number and nature of the common economic 
factors affecting the time evolution of logarithm of price of well defined 
groups of stocks.
\end{abstract}

\bigskip
\bigskip

Financial markets are well defined complex systems. They are studied by 
economists, mathematicians and, recently, also by physicists. The paradigm 
of mathematical finance is that the time series of stock returns are
unpredictable \cite{Samuelson65}. Within this paradigm, time evolution 
of stock returns are well
described by random processes. A key point is if the random processes of
stock returns time series of different stocks are uncorrelated or, conversely, 
if common economic factors are present in financial markets and are driving 
several stocks at the same time. Common economic factors were 
originally introduced by Ross in his arbitrage pricing theory model
\cite{Ross76}. Since its introduction, the problem of the number and 
nature of common economic factors has been a considerable controversial
issue. On the side of modeling of financial markets
by using tools and procedures developed to model physical systems
\cite{Mandelbrot63,Kadanoff71,Mantegna91,Bak93,Bouchaud94,Mantegna95,Ghashghaie96,BakPA}, there is the need to characterize the topological arrangement of different 
stocks traded in a financial markets. Similar information is essential 
to attempt to model financial markets in terms of nonlinear 
systems in the presence of external and/or quenched noise.
 
The motivation of the present study is twofold. The first motivation
concerns the search for the kind of topological arrangement which is 
present between the stocks of a portfolio traded in a financial market.
The second motivation is the 
search of empirical evidence about the existence and nature of common 
economic factors which drive the time evolution of
stock prices. 

In the present analysis, I investigate the hierarchical structure present 
in a portfolio of $n$ stocks traded in a financial market. The observable 
which is used to detect the topological arrangement of the stocks of
a given portfolio is the synchronous correlation coefficient of the daily 
difference of logarithm of closure price of stocks. The 
correlation coefficient is computed between all the possible pairs of stocks 
present in the portfolio in a given time period. In this letter, 
I report results obtained by investigating the portfolio of the stocks used to
compute the Dow Jones industrial average index and the portfolio of
stocks used to compute the Standard and Poor's 500 index in the time 
period from July 1989 to October 1995. Both indices mainly
describe the performance of the New York Stock Exchange.

The starting point of my investigation is to quantify the degree of 
similarity between the synchronous time evolution of a pair of stock 
price by the correlation coefficient \cite{Feller71}
\begin{equation}
\rho_{ij}=\frac{<Y_i Y_j>-<Y_i><Y_j>}{\sqrt{(<Y_i^2>-<Y_i>^2)(<Y_j^2>-<Y_j>^2)}}
\end{equation}
where $i$ and $j$ are the numerical labels of stocks, $Y_i=\ln P_i(t)-\ln P_i(t-
1)$ and $P_i(t)$ is the closure price of the stock 
$i$ at the day $t$. The statistical average is a temporal average performed
on all the trading days of the investigated time period.

For both portfolios, I determine the $n \times n$ matrix of correlation 
coefficients for daily logarithm price differences (which almost 
coincides with returns). By definition, $\rho_{ij}$ can vary from -1 (completely
anti-correlated pair of stocks) to 1 (completely correlated pair of stocks).
When $\rho_{ij}=0$ the two stocks are uncorrelated. 
 
The matrix of correlation coefficients is a symmetric matrix with 
$\rho_{ii}=1$ in the main diagonal. Hence, in each portfolio, $n~(n-1)/2$ 
correlation coefficients characterize the matrix completely.
An investigation of the statistical properties of the set of correlation
coefficients is published elsewhere \cite{Mantegna97}.
In this letter, I investigate the correlation coefficient matrix to 
detect the hierarchical organization present inside the stock market.
In the search for an appropriate topological arrangement of stocks
of a given portfolio, I first look for a metric. The correlation
coefficient of a pair of stocks cannot be used as a distance between 
the two stocks because it does not fulfill the three axioms that define 
an Euclidean metric. However a generalized metric can be defined  
using as distance an appropriate function of the correlation coefficient.
The chosen function is 
\begin{equation}
d(i,j)=1-\rho_{ij}^2 .
\end{equation}

\noindent
With this choice $d(i,j)$ fulfills the three axioms of an Euclidean metric --
(i) $d(i,j)=0$ if and only if $i=j$; (ii) $d(i,j)=d(j,i)$ and (iii)
$d(i,j) \le d(i,k) +d(k,j)$, for all practical purposes. The first axiom is 
valid because $d(i,j)=0$ if and only if the
correlation (or the anticorrelation) is complete $(|\rho|=1$, namely only 
if the two stocks perform the same stochastic process or stochastic processes
which are mirror images the one of the other). The second axiom is valid
because the correlation coefficient matrix and hence the distance matrix 
$\bf{D}$ is symmetric by definition, and the third axiom has been verified 
numerically in all portfolios investigated.

The distance matrix $\bf{D}$ is then used to determine the minimal spanning tree 
\cite{Papa82} connecting the $n$ stocks of the portfolio. The minimal spanning 
tree (MST) is attractive because provides an arrangement of stocks which selects the
most relevant connections of each point of the set. Moreover 
the minimal spanning tree gives, in a direct way, the subdominant ultrametric 
hierarchical organization of the points (stocks) of the investigated portfolio.
In the rest of this letter, I will show that the hierarchical organization found 
through the minimal spanning tree associated with the distance matrix $\bf{D}$ 
is of great interest from an economic point of view. In particular, by assuming
this kind of topology, I am able to isolate groups of stocks which makes sense from
an economic point of view by starting from the information carried by the 
time series of prices only.
The classification of the groups of stocks obtained with my analysis
of the correlation coefficients  
is performed by using the industry and subindustry sectors reported in the Forbes 49th
annual report on American industry.

For each portfolio investigated, starting from the distance matrix ${\bf D}$, it
is possible to obtain the minimal spanning tree (MST) connecting the stocks of 
the portfolio. 
In fig. 1a, I show the minimal spanning tree for the Dow Jones industrial
average portfolio of stocks. Each circle represents a stock, labeled by its
tic symbol (for example KO is Coca Cola Com. , PG is Procter \& Gamble,
etc. See the caption of the figure for more details).The connections between
stocks are shown by segments and the color of these segments is related to 
the distance between stocks. In fig. 1b the hierarchical tree of the
subdominant ultrametric \cite{Rammal86} associated to the MST is shown.
An inspection of the MST and of the associated hierarchical tree show the 
existence of three groups of stocks.
The observed grouping has a direct economic explanation.
The more evident and strongly connected group is the group of stocks
CHV, TX and XON namely Chevron , Texaco and Exxon. These three companies are
working in the same industry (energy) and in the same subindustry (international 
oils). A second group is formed by AA and IP, namely by
Alcoa (working in the subindustry sector of nonferrous metals) and International 
Paper (working in the subindustry sector of paper and lumber).
Both companies provide raw materials. The third group involves companies which 
are in industry sectors which deals with consumer nondurables (Procter \& 
Gamble, PG) and food drink and tobacco (Coca Cola, KO).

The same investigation is repeated for the set of stocks used to compute
the Standard and Poor's 500 index. In this case the larger size of the
portfolio allows to perform a more refined test of the detected hierarchical 
structure of stocks.
In my analysis, I considered only the companies which were present in 
the S\&P 500 index for the entire period investigated. With this constrain the 
portfolio is composed of
443 stocks. Due to the size of the portfolio investigated, the obtained 
minimal spanning tree cannot be shown in a single figure in a legible way. In 
Fig. 2,
I show some of the parts of the MST which are strongly connected. A group of 
financial
services, capital goods, retailing, food drink \& tobacco and consumer
nondurables companies is shown in I;  the group of oil companies is 
shown in II , while
III is the group  of communication and electrical utility companies
and IV is the group of raw material companies. The portfolio
of stocks used to compute the S\&P 500 index is characterized by 
a hierarchical structure of stocks which is much more detailed than the one
observed in the case 
of the DJIA portfolio.
The structure of the minimal spanning tree of the portfolio of stocks of the 
S\&P 500 index, shows many groups of stocks which are homogeneous 
from an economic point of view. 
A detailed inspection of the hierarchical tree associated
to the MST provides a large amount of economic information. It is impossible to 
put in a single legible figure the 
complete hierarchical tree of a so broad portfolio. In fig. 3, we then show only 
the branching of the tree up to the level of homogeneous
groups. This means that lines in the hierarchical tree shown in 
fig. 3 are always ending in a group of stocks which contains at least 2 stocks
(but usually more). The branches of single stocks departing from the 
tree are not shown to make the figure readable. In the caption of fig. 3,
I give full details about the
stocks belonging to the groups shown in the figure, together with their 
classification in industry sectors and/or subsectors. With only a few exceptions 
the groups are homogeneous with respect to industry and often also subindustry 
sectors suggesting that set of stocks working in the same industry
and subindustry sectors respond, in a statistical
way, to the same economic common factors.
In some cases, my analysis, based on the statistical analysis of correlation
coefficients between pairs of stock returns, refines the classification in 
sectors and subsectors used by Forbes. For example, ores, aluminum and copper 
are all classified metals as industry and nonferrous metals as subindustry. From 
my analysis, I detect that they respond to quite
different common economic factors. Specifically, ores companies are grouped 
in a cluster 
which is the most distant form all the others groups of stocks of the tree, while 
aluminum and copper companies constitute a subgroup of the group containing raw 
materials companies.

The detection of a hierarchical structure in a broad portfolio of stocks
traded in a financial market is consistent with the assumption that the 
time series of returns of a stock is affected by a number
of economic factors. The analysis shows that the number and the relative influence 
of these factors is specific to each stock. In general, stocks or groups of 
stocks that depart early 
from the tree (at high values of the distance $d(i,j)$) are mainly 
controlled by economic factors which are specific to the considered group
(for example gold price for the stocks of the group 1 of the tree (see Fig. 3)
which is composed only by companies involved in gold mining). When departure occurs for 
(moderately) low values of $d$, the stocks are affected either by economic
factors which are common to all stocks and by other economic factors
which are specific to the considered set of stocks. The relative relevance of 
these factors is quantified by the length of the segment (or segments) observed 
for each group from one branching to the successive one.

In conclusion, the present study shows that the MST and the associated
subdominant ultrametric hierarchical tree, obtained starting from the distance
matrix of Eq. (2), selects a topological space for the stocks of a 
portfolio traded in a financial market which is able to give 
an economic meaningful taxonomy. This topology is useful in the theoretical
description of financial markets and in the search of economic common factors 
affecting specific groups of stocks. The topology and the hierarchical structure 
associated to it, is obtained by using information present in the time series of stock 
prices only. This result shows that time series of stock prices are carrying 
valuable (and detectable) economic information.
 
\acknowledgments
I thank INFM and MURST for financial support.

\begin{figure}
\caption{(a) Minimal spanning tree connecting the 30 stocks used to 
compute the Dow Jones Industrial Average. The distance matrix ${\bf D}$ is
obtained starting from the correlation coefficients measured between all pairs 
of stocks in the portfolio in the
time period from July 89 to October 95. The 30 stocks are labeled by their
tic symbols (AA--Alcoa, ALD--Allied Signal, AXP--American Express Co,
BA--Boeing Co, BS--Bethlehem Steel, CAT--Caterpillar Inc., CHV--Chevron Corp.,
DD--Du Pont, DIS--Walt Disney Co., EK--Eastman Kodak Co., GE--General Electric,
GM--General Motors, GT--Goodyear Tire, IBM--IBM Corp., IP--International Paper,
JPM--Morgan JP, KO--Coca Cola Co., MCD--McDonalds's Corp., MMM--Minnesota 
Mining, MO--Philips Morris, MRK--Merck \& Co Inc., PG--Procter \& Gamble,
S--Sears Roebuck, T--AT\&T, TX--Texaco Inc., UK--Union Carbide, UTX--United 
Tech, WX--Westinghouse, XON--Exxon Corp. and Z--Woolworth). The colors of 
segments connecting stocks are proportional to the distance between the stocks
(yellow $0.65<d(i,j)\le 0.70$, green $0.70<d(i,j)\le 0.75$, turquoise 
$0.75<d(i,j)\le0.80$, cyan $0.80<d(i,j)\le 0.85$, blue $0.85<d(i,j)\le 0.90$ and 
violet $0.90<d(i,j)\le 0.95$).
(b) Hierarchical tree of the subdominant ultrametric associated with the
minimal spanning tree of a). In the tree, groups of stocks, homogeneous with 
respect to the economic activities of the companies are detected: (i) oil
companies (Exxon, Texaco  and Chevron); (ii) raw material companies (Alcoa and
International paper) and (iii) companies working in the sectors of consumer
nondurable products (Procter \& Gamble) and food and drinks (Coca Cola). The
distance at which individual stocks are branching from the tree is given by the 
$y$ axis.}
\label{fig1}
\end{figure}
 
\begin{figure}
\caption{Partial regions of the minimal spanning tree of the portfolio of stocks 
used to compute the S\&P 500 index. 
The distance matrix ${\bf D}$ is
obtained starting from the correlation coefficients measured between all pairs 
of stocks in the portfolio traded in the
time period from July 89 to October 95. The four panels show four strongly 
connected large groups of stocks observed for $d(i,j) \le 0.81$.
Circles represents stocks which are labeled by their stock exchange tic 
symbols. The colors of segments joining circles are proportional to the 
distance between stocks (magenta $0.5<d(i,j)\le 0.55$, red $0.55<d(i,j)\le 0.6$, 
orange $0.6<d(i,j)\le0.65$ and other colors as in fig. 1a).
The dashed segments are connections between stocks which join different
regions of the minimal spanning tree having distances $d(i,j) > 0.81$.
In panel I, the group of financial service companies (AHM, BAC, BBI, BK, 
BKB, BT,  CCI, CHL, CMB, FFB, FNB, FNM, FTU, GDW, GWF, I, JPM, MER, ONE, 
PNC, SNC and WFC), capital
goods companies (EMR and GE), retailing companies (HD and WMT), consumer 
nondurables
companies (CL and PG) and food and drinks companies (KO and PEP) are shown.
Du Pont company (DD) is joining this group of stocks to the group of oil
companies of panel II. Panel II shows international oil companies (AN, CHV,
MOB, RD, TX and XON), other energy companies (AHC, ARC, KMG, P and UCL) and 
oilfield service companies (BHI, DI, HAL and SLB).
UN (Unilever) is the only stock of the group which is not directly 
homogeneous to industry subsectors of international oil and other energy.
Panel III shows 
companies which are working in the industry sector of electric utilities
(AEP, BGE, CPL, CSR, D, DTE, DUK, ED, FPL, NSP, OEC, PCG, PE, PEG and SO) and
in the subindustry sector of telecommunications (AIT, BEL, BLS, GTE, NYN, SBC
and USW). T, namely AT\&T Corp., connects this group of stocks to other groups
through General Electric (GE). Last panel (IV) shows a group of companies 
with activities in the sectors of raw materials. Two subgroups are observed:
(i) companies working in the industry sector of metals and subindustry 
nonferrous materials (AA, AL, AR, PD, N and RLM) and companies working in the
industry sector of forest products and packaging (CHA, IP, LPX, MEA, TIN,
UCC and WY). The two subgroups are connected between them through the link
between Alcoa (AA) and International paper (IP). They are connected to
other groups through Minn Mining \& MFG (MMM) and General Electric.   
}
\label{fig2}
\end{figure}
 
Fig. 3 Main structure of the minimal spanning tree of the
portfolio of stocks used to compute the S\&P 500 index. Each line 
ending in the bottom corresponds to a group of stocks composed by 
at least two stocks. Lines are ending when the first bifurcation 
inside the group is observed. Individual stocks departing from the 
main tree are not shown for the sake of clarity. Groups are labeled 
with integers ranging from 1 to 44. The branching of each group from
the main tree and inside the group are occurring at a distance given 
by the $d(i,j)$ scale. Below, I report for each group detected in the
MST the observed common industry sector (IS) and subindustry sector 
(SS) together with the tic symbols of the stocks belonging to the group.
The IS and SS of stocks are the ones used in the 49th Forbes annual report 
of American industry (accessible on the web at the address www.forbes.com).
1. IS metals, SS nonferrous metals (gold), stocks ABX, ECO, HM,
NEM and PDG; 2. IS construction, SS residential builders, stocks CTX,
KBH and PHM; 3. no common industry sector, stocks ACK and MAS; 4.
IS travel and transport, SS trucking and shipping, stocks ROAD and YELL;
5. IS consumer nondurables, SS photography and toys, stocks HAS and MAT;
6. no common industry sector, stocks CBS and CCB; 7. IS metals, SS steel,
stocks BS, IAD and X; 8. IS consumer durables, SS automotive parts, stocks
DCN and ETN; 9. IS travel and transport, SS airlines, stocks AMR, DAL, LUV 
and U; 10. IS entertainment and information, SS broadcasting and cable,
stocks CMCSA and TCOMA; 11. financial services, SS lease and finance, 
stocks BNL and HI; 12. IS energy, SS oilfield services, stocks BHI, DI, HAL
and SLB; 13. IS energy, SS international oils, stocks AN, CHV, MOB, TX, XON
and ARC (IS other energy); 14. no common industry sector, stocks RD and UN;
15. IS capital goods, SS heavy equipment, stocks CAT, DE, IR and TEN (IS
forest products and packaging); 16. IS business services and supplies, SS
environmental and waste, stocks BFI and WMX; 17. IS construction, SS commercial
builders, stocks FLR and FWC; 18. IS consumer durables, SS automobiles and 
trucks, stocks C, F and GM; 19. IS food drink and tobacco, SS tobacco, 
stocks AMB and MO; 20. IS entertainment and information, SS publishing,
stocks GCI, KRI and TRB; 21. IS forest products and packaging, SS
paper and lumber, stocks BCC, CHA, IP, LPX, MEA, UCC, WY and TIN (SS
packaging); 22. IS metals, SS nonferrous materials, stocks AR and PD; 
23. IS metals, SS nonferrous materials, stocks AL and N;
24. IS metals, SS nonferrous materials, stocks AA and RLM;
25. IS computer and communications, SS peripherals \& equipment or
software, stocks AMAT, CPQ, HWP, INTC, MOT, MSFT, NOVL, NSM and TXN;
26. IS Electric utilities, SS regional area, stocks  AEP, BGE, CPL, CSR, D, DTE, 
DUK, ED, FPL, NSP, OEC, PCG, PE, PEG and SO; 27. IS computer and communications, 
SS telecommunications, stocks AIT, BEL, BLS, GTE, NYN, SBC
and USW; 28. IS retailing, SS department stores and drug \& discount,
stocks DH, DDS and MAY; 29. no common industry sector, stocks DD, DOW and 
VO; 30. IS travel and transport, SS railroads, stocks BNI, CRR, CSX, NSC 
and UNP; 31. IS food drink and tobacco, SS food processors, stocks CPC, 
GIS, K and SLE; 32. no common industry sector, stocks AET, and CI;
33. IS Insurance, SSs property \& casualty and diversified, stocks
AIG, CB and GRN; 34. IS health, SS drugs, stocks PFE and SGP; 35. IS 
health, SS drugs, stocks BMY and MRK; 36. IS consumer nondurables, SS
personal products, stocks CL and PG; 37. IS food drink and tobacco, 
SS beverages, stocks KO and PEP; 38. IS retailing, no common SS, stocks
HD and WMT; 39. IS capital goods, SS electrical equipment, 
stocks EMR and GE; 40. IS financial services, no common SS, stocks 
FNM and GDW; 41. IS financial services, SS thrift institutions, stocks 
AHM and GWF; 42. IS financial services, SS multinational banks, stocks 
BT and JPM; 43. IS financial services, SS regional banks, stocks I (no more 
traded) and WFC; 44. IS financial services, SS multinational banks, stocks 
BAC, CCI, CHL (no more traded) and CMB.          

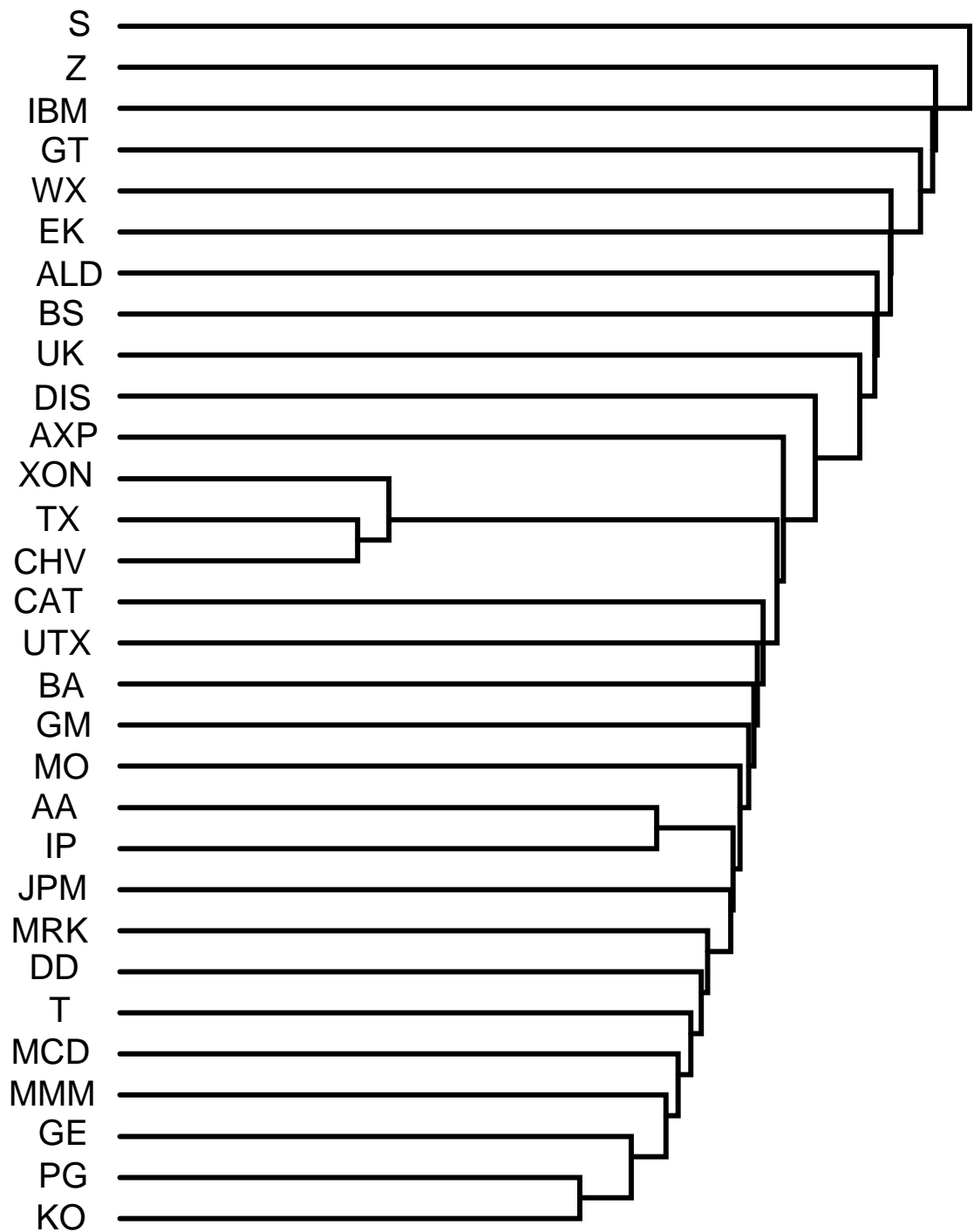
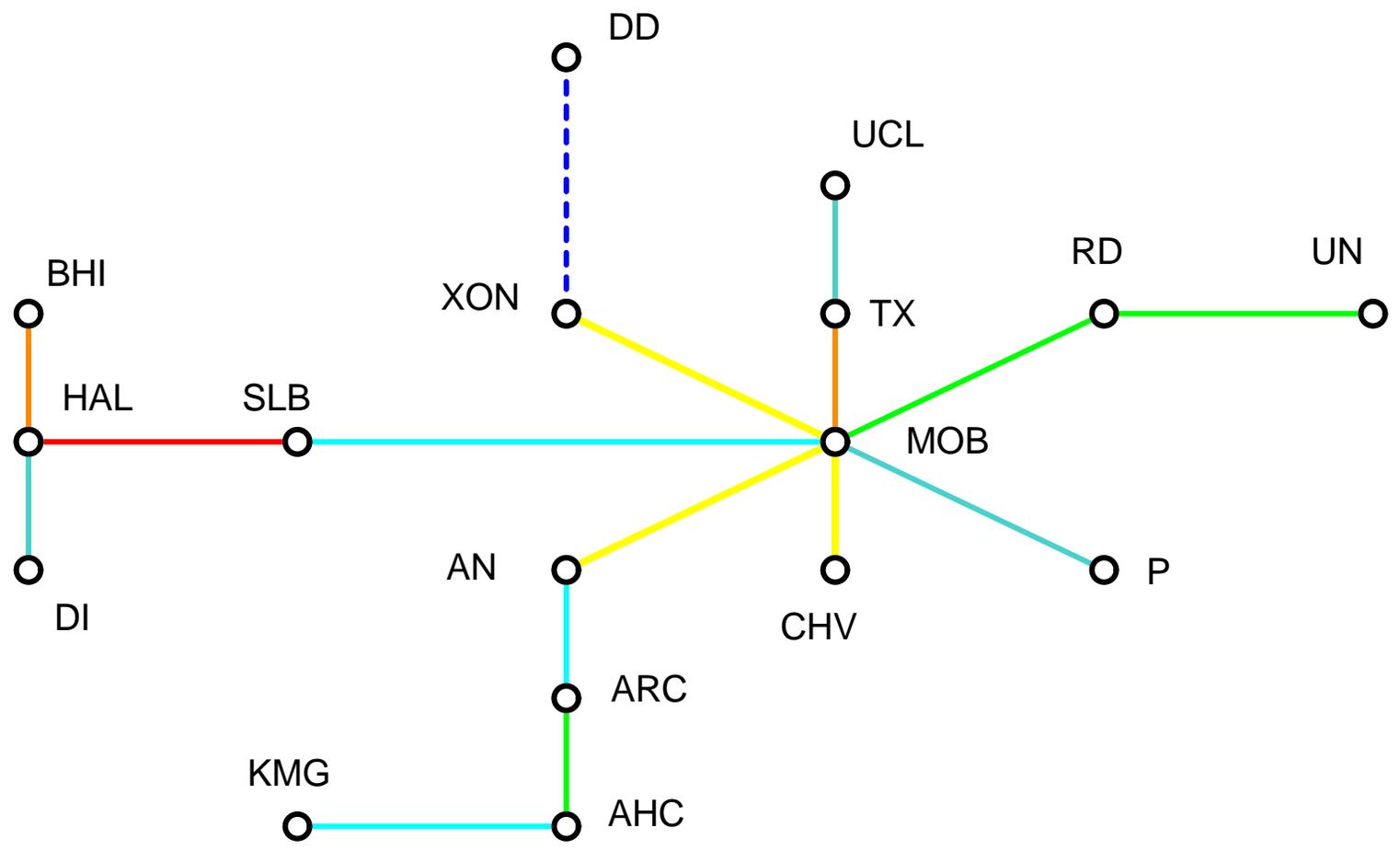
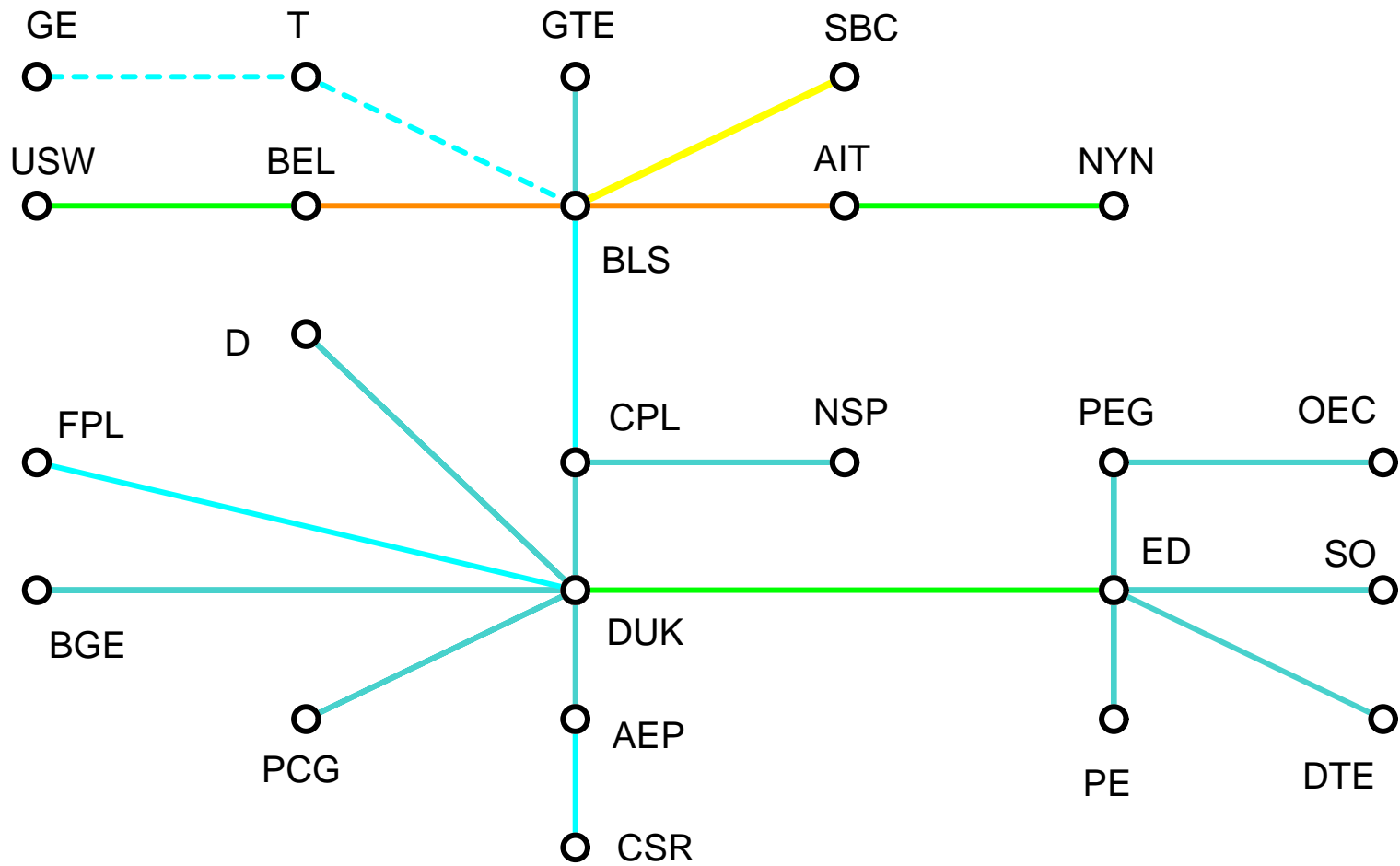
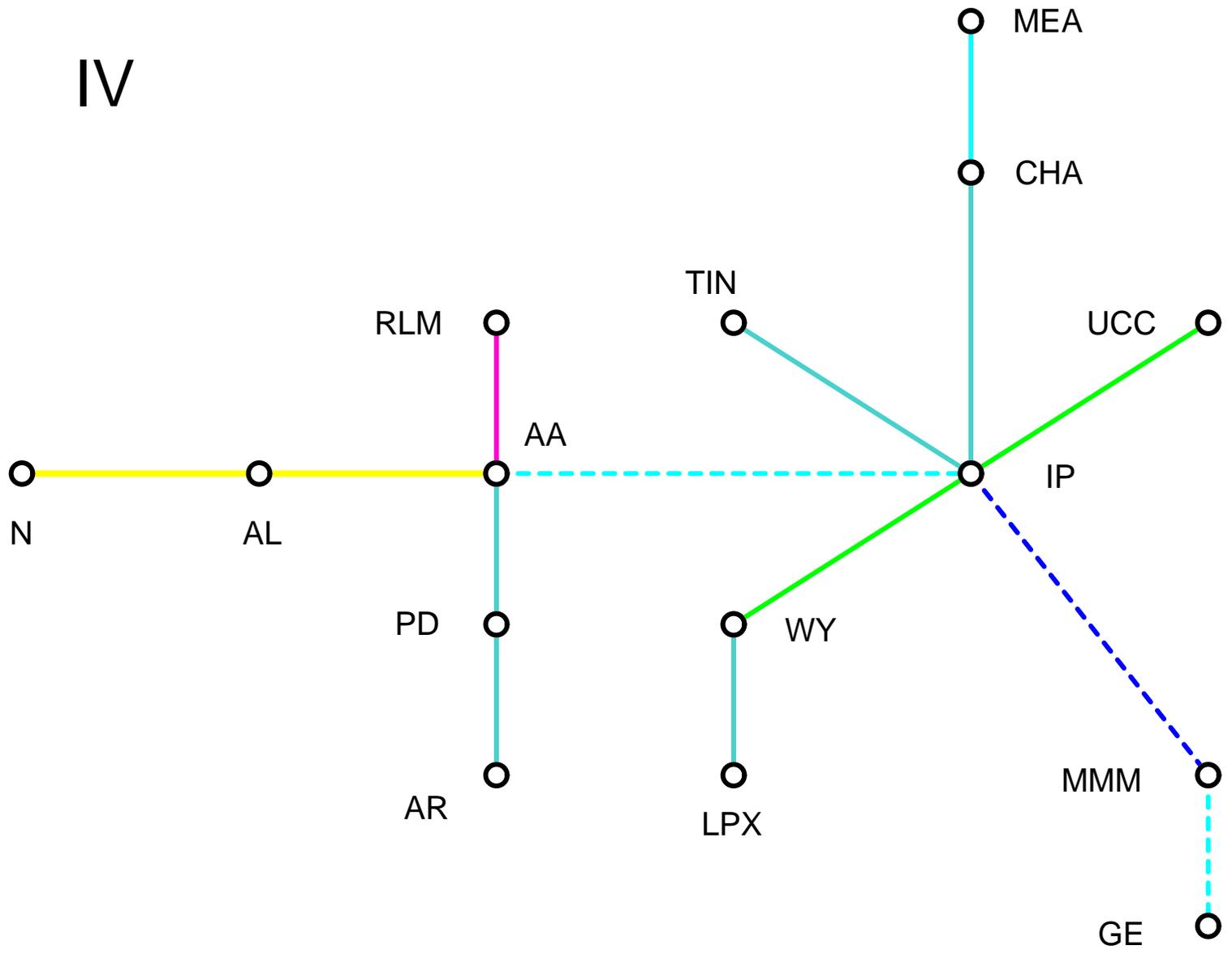
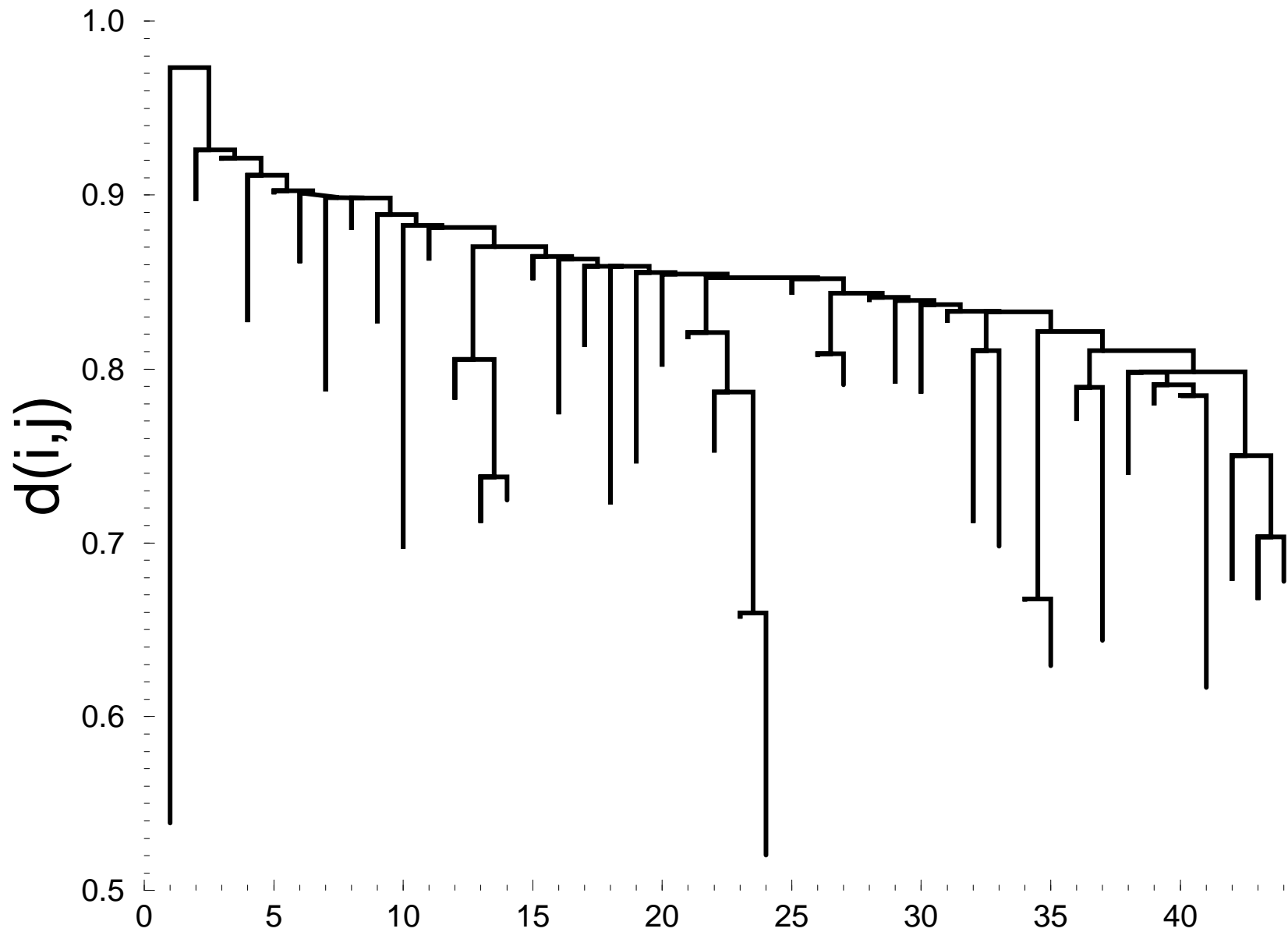
\end{document}